\documentclass[twocolumn,aps,showpacs,prb,superscriptaddress]{revtex4-1}
\usepackage{natbib}
\usepackage{url}
\usepackage{bm}
\usepackage{textcase}
\usepackage{color}
\usepackage{graphicx}
\usepackage{amsmath}
\usepackage{ulem}
\usepackage[colorlinks=true,citecolor=blue]{hyperref} 
\hypersetup{pdftitle={Density-Matrix Functionals from Green's functions}}
\hypersetup{pdfauthor={P.E. Bloechl, Th. Pruschke, M. Potthoff}}
\hypersetup{pdfdisplaydoctitle}
\newcommand{\Tr}{{\rm Tr}}
\newcommand{\mat}[1]{{\boldsymbol #1}}
\newcommand{\e}[1]{{\rm e}^{#1}}
\newcommand{\mathrho}{{\bar{\mat{h}}}}
\newcommand{\hathrho}{{\hat{\bar{h}}}}
\newcommand{\matGrho}{{\bar{\mat{G}}}}
\DeclareMathOperator*{\stat}{stat}

\begin{document}
\title{Density-matrix functionals from Green's functions}
\author{Peter E. Bl\"ochl}
\email[corresponding author: ]{peter.bloechl@tu-clausthal.de}
\affiliation{Clausthal University of Technology, Institute for
Theoretical Physics, Leibnizstr.10, D-38678 Clausthal-Zellerfeld,
Germany}
\affiliation{Georg-August Universit\"at G\"ottingen, Institute for
Materials Physics, Friedrich-Hund-Platz 1, D-37077 G\"ottingen,
Germany}
\author{Thomas Pruschke} 
\affiliation{Georg-August Universit\"at G\"ottingen, Institute for
Theoretical Physics, Friedrich-Hund-Platz 1, D-37077 G\"ottingen,
Germany}
\author{Michael Potthoff\;}
\affiliation{I. Institute for Theoretical Physics, 
University of  Hamburg, 
Jungiusstr. 9, D-20355 Hamburg, Germany}
\date{\today} 
\begin{abstract}
The exact reduced density-matrix functional is derived from the
Luttinger-Ward functional of the single-particle Green's function.
Thereby, a formal link is provided between diagrammatic many-body
approaches using Green's functions on the one hand and theories based
on many-body wave functions on the other.  This link can be used to
explicitly construct approximations for the density-matrix functional
that are equivalent to standard diagrammatic re-summation techniques
and to non-perturbative dynamical mean-field theory in particular.
Contrary to functionals of the Green's-function, the exact
density-matrix functional is convex and thus provides a true minimum
principle which facilitates the calculation of the grand potential and
derived equilibrium properties.  The benefits of the proposed
Green's-function-based density-matrix functional theory for
geometrical structure optimization of strongly correlated materials
are discussed.
\end{abstract}
%
\maketitle 
\section{Introduction}

Consistent and reliable approximations for extended systems of
correlated electrons are ideally generated by making use of general
variational principles.  Two fundamentally different concepts have
been pursued in the past, namely wave-function-based and
Green's-function-based methods.

(i) Wave-function-based methods make use of the Ritz variational
principle, which relies on the optimization of the many-body
ground-state wave function or, at finite temperatures, the statistical
operator.  Density-functional theory (DFT)
\cite{hohenberg64_pr136_B864, kohn65_pr140_1133, levy79_pnas76_6062,
  dreizler90_book} and reduced density-matrix functional theory
(rDMFT) \cite{gilbert75_prb12_2111, levy79_pnas76_6062,
  coleman63_rmp35_668, mueller84_pl105A_446, baerends01_prl87_133004,
  sharma08_prb78_201103, lathiotakis09_pra79_40501} are based on the
optimization of the electron density or the reduced one-particle
density matrix, respectively, and the underlying variational
principles derive from the Ritz principle.  The Ritz principle
actually represents a \textit{minimum} principle, i.e., the
ground-state energy or, at finite $T$, \cite{mermin65_pr137_A1441} the
grand potential is at a minimum for the exact (``physical'') wave
function or density operator.  Analogously, within DFT and rDMFT, the
physical electron density or the physical one-particle reduced density
matrix, respectively, \textit{minimize} the corresponding functional.
Within condensed-matter electronic-structure theory, approximations
derived from those wave-function-based approaches, e.g.\ the
local-density approximation
\cite{hohenberg64_pr136_B864,kohn65_pr140_1133} within DFT, are
usually employed to study weakly or moderately correlated materials.

(ii) Strongly correlated electron systems are often studied with
Green's-function techniques.  The one-electron Green's function is
closely related to the spectrum of one-particle excitations accessible
to photoemission spectroscopy.  It can be obtained from a
\textit{dynamical} variational principle,
\cite{luttinger60_pr118_1417,baym61_pr124_287,baym62_pr127_1391} named
after Kadanoff and Baym, which expresses the grand potential as a
functional of the frequency-dependent Green's function or as a
functional of the Green's function and the self energy.  This and
related variational principles
\cite{chitra00_prb62_12715,potthoff03_epjb32_429,georges04_aipcp715_3}
are based on the Luttinger-Ward
functional.\cite{luttinger60_pr118_1417,potthoff06_condensmatterphys9_557}
Dynamical variational principles represent \textit{stationary}
principles, i.e., the grand potential is stationary -- but not
necessarily extremal -- at the physical Green's function and the
physical self energy.  Despite some
attempts\cite{chitra01_prb63_115110,nevidomskyy08_prb77_75105}, there
is, with the exception of certain highly symmetric cases
\cite{kotliar99_epjb11_27}, no known general dynamical functional
which is convex.  Self-energy functional theory
\cite{potthoff03_epjb32_429} has explicitly shown that evaluating a
dynamical functional at trial self-energies away from a stationary
point may yield a grand potential smaller than the physical one.

The reason for this deficiency of Green's-function-based theories
lies in the fact that there is no general relation by which dynamical
functionals of the Green's function or self energy can be transformed
into functionals of the many-body wave function while preserving the
variational properties.  

In this paper, we derive the density-matrix functional from the
Luttinger-Ward functional and thus provide a link between
wave-function-based and Green's-function-based approaches.  
The density-matrix functional for a given one-particle density matrix
is obtained as the value of a functional of the one-particle Green's
function and the self energy at its stationary point.  The
density-matrix functional obtained in this way, inherits a true
minimum principle from its wave-function-based origin.

The conceptual benefit from this construction is that well-known
diagrammatic approximations within many-body perturbation theory will
be ``translated'' into corresponding approximations within the
framework of rDMFT.  Standard \textit{conserving approximations}
\cite{baym61_pr124_287,baym62_pr127_1391} include the Hartree-Fock
(HF) approximation and weak-coupling approximations, such as the
random-phase approximation (RPA), $GW$ or the fluctuation-exchange
(FLEX) approximation. However, also the non-perturbative dynamical
mean-field theory can be obtained variationally using the one-particle
density matrix as the basic variable.

The stationary property of dynamical functionals has been
exploited earlier, namely to obtain improved approximations for the
total energy using an approximate Green's function or self
energy.\cite{almbladh99_ijmpb13_535,dahlen03_jcp120_6826,
  dahlen04_prb69_195102, dahlen05_ijqc101_512, dahlen06_pra73_12511}.
With a minimum principle at hand, one is in a position to compute
strict upper bounds to the total energy additionally and thereby to
compare the quality of different approximate Green's functions.

Convexity of the density-matrix functional is also very helpful for
numerical purposes: \cite{press07_book} There are a number of efficient
general techniques for finding an extremum of a multidimensional
function.  For the search of stationary points, including saddle
points, on the other hand, one generally has to resort to less
efficient multidimensional root finding.

Furthermore, geometry optimization is greatly facilitated if there is
a minimum principle available for the total energy given in terms of
electronic degrees of freedom.  This is of interest for materials with
strong electronic correlations, which often exhibit a strong mutual
dependence between the electronic equilibrium and the geometrical
structure. In particular, electronically driven phase transitions are
often accompanied by structural relaxations.  Here, the prototypical
method that we have in mind is the fictitious-Lagrangian approach of
\textit{ab initio} molecular dynamics,\cite{car85_prl55_2471} which performs
the electronic and geometric optimization simultaneously.  One big
advantage of this method is that the calculations of forces do not
require the inclusion of the linear response of the electronic
structure to a virtual displacement, even if the electronic structure
is not at its ground state.  \textit{Ab initio} molecular dynamics is
not only useful for structure optimization, but also for the study of
atomic motion either for investigating the dynamics or, exploiting the
ergodic principle, the ensemble properties at finite
temperature. However, \textit{ab initio} molecular dynamics depends on a true
minimum principle for the electronic degrees of freedom.

Finally, a further benefit of constructing density-matrix
functionals using Green's function techniques consists in the
possibility to find \textit{explicit} expressions for approximate
density-matrix functionals that are consistent with specific
diagrammatic approximations or have a comparable quality or
reliability.  This would allow us to bypass the complexity of a
diagrammatic many-body calculation, i.e., to work with the much
simpler density matrix rather than with Green's functions or
self-energies.  For geometry optimization, but also in other contexts,
this would be extremely helpful.  Although dynamical
(frequency-dependent) quantities will be disregarded in such an
approach, one would still have full access to spectral properties in
the end.  Namely, the complexity of calculating spectral properties
can be postponed to the converged state.  Quite generally, we will
show how to obtain a spectral function from a \textit{converged} density
matrix consistent with the approximation underlying the density-matrix
calculation.

The purpose of the present paper is to present the theoretical setup.
Sec.~\ref{sec:basics} presents the notation and introduces the
basics of density-matrix functional theory as well as of the
Luttinger-Ward functional.  In Sec.~\ref{sec:kbtordmft}, we show
how an intrinsic symmetry of the Kadanoff-Baym functional leads to the
definition of the density-matrix functional in terms of a stationary
functional of Green's functions and self energies. In
Sec.~\ref{sec:approximations}, we discuss some
common approximations and in
Sec.~\ref{sec:forces}, we discuss the implications on force
calculations and spectral properties.

\section{Variational principles with wave functions and Green's functions}
\label{sec:basics}
\subsection{Density-matrix functional approach}
The grand potential
for a many-particle system has the form
\begin{eqnarray}
\Omega_{\beta,\mu}(\hat{h}+\hat{W})=-\frac{1}{\beta}\ln\Bigl[\Tr\Bigl\lbrace
\e{-\beta(\hat{h}+\hat{W}-\mu\hat{N})}\Bigr\rbrace\Bigr]
\;,
\end{eqnarray}
where $\beta=1/(k_BT)$ , with the Boltzmann constant $k_B$ and the
temperature $T$. The trace is performed over the 
Fock space of the electron gas. Furthermore,
\begin{eqnarray}
\hat{h}=\sum_{a,b}h_{a,b}\hat{c}_a^\dagger\hat{c}_b
\label{eq:hath}
\end{eqnarray}
is the non-interacting part of the Hamiltonian expressed by creation
and annihilation operators in an orthonormal one-particle basisset, 
\begin{eqnarray}
\hat{W}=\frac{1}{2}
\sum_{a,b,c,d}U_{a,b,d,c}\hat{c}_a^\dagger\hat{c}_b^\dagger
\hat{c}_c\hat{c}_d
\end{eqnarray}
is the electron-electron interaction, and $\hat{N}$ is the particle number
operator
\begin{eqnarray}
\hat{N}=\sum_{a} \hat{c}_a^\dagger\hat{c}_a
\;.
\end{eqnarray}

In the following, bold-faced symbols are matrices in the 
one-particle Hilbert space. Derivatives with respect to matrices are
interpreted in the form $\Bigr(\frac{\partial
  Y}{\partial\mat{A}}\Bigr)_{a,b} =\frac{\partial
  Y}{\partial{A}_{b,a}}$. 
  
The wave-function approach to the grand-canonical ensemble is to
minimize the grand potential, expressed as a functional of (fermionic)
many-particle wave functions $|\Phi_j\rangle$ in Fock space and their
probabilities $P_j$:
\begin{eqnarray}
\Omega_{\beta,\mu}(\hat{h}+\hat{W})&=&
\min_{P_j\ge0,|\Phi_j\rangle}\stat_{\mat{\Lambda},\lambda}
\biggl\lbrace
\frac{1}{\beta}\sum_j P_j\ln[P_j]
\nonumber\\
&&+\sum_jP_j\langle\Phi_j|\hat{h}+\hat{W}-\mu\hat{N}|\Phi_j\rangle
\nonumber\\
&&-\sum_{i,j}\Lambda_{i,j}\Bigl(\langle\Phi_j|\Phi_i\rangle-\delta_{j,i}\Bigr)
\nonumber\\
&&-\lambda\Bigl(\sum_j P_j-1\Bigr)
\biggr\rbrace
\;.
\end{eqnarray}
With $\mu$, we denote the chemical potential of the electrons.  We
introduce the symbol ``$\stat$'' to denote a stationary condition,
which may be an extremum or a saddle point.

The constraints of orthonormal wave functions and the sum rule for the
probabilities are enforced using 
Lagrange multipliers
$\Lambda_{i,j}$ and
$\lambda$.  The probabilities must be positive. In practice, this
requirement is enforced by expressing the probabilities as squares of
real-valued variables.

Following the constrained-search method of
Levy\cite{levy79_pnas76_6062}, the minimization with respect to all
many-particle wave functions can be divided into two steps. (1) For
each one-particle reduced density matrix $\mat{\rho}$, we
collect all fermionic many-particle ensembles, consisting of sets of
antisymmetric many-particle wave functions $|\Phi_j\rangle$ in Fock
space and their probabilities $P_j$, which yield this density
matrix via
\begin{eqnarray}
\rho_{a,b}:=\sum_j
P_j\langle\Phi_j|\hat{c}^\dagger_b\hat{c}_a|\Phi_j\rangle\;.
\label{eq:defden}
\end{eqnarray}
Then, one identifies the minimum of the grand potential for
this subset. (2) In the second step, we minimize the grand
potential with respect to the one-particle density matrix. 

The first part in this two-step minimization defines a functional of
the one-particle reduced density matrix. This density-matrix
functional is then used to determine the minimum in the second part,
which is performed without referring to many-particle wave functions:
All relevant information from the many-particle wave functions has
been encoded in the density-matrix functional.

To obtain a density-matrix functional, one splits the grand
potential into contributions that can be expressed by the one-particle
density matrix alone and terms that can only be obtained from the
many-particle wave function. Thus, we obtain
\begin{eqnarray}
\Omega_{\beta,\mu}(\hat{h}+\hat{W})&=&
\min_{\mat{\rho}}\Bigl\lbrace\Tr[\mat{\rho}\left(\mat{h}-\mu\mat{1}\right)]
+F^{\hat{W}}_\beta[\mat{\rho}]
\Bigr\rbrace
\label{eq:rdmftgrandcanonicalpotouter}
\end{eqnarray}
with
\begin{eqnarray}
F^{\hat{W}}_\beta[\mat{\rho}]&=&
\min_{P_j\ge0,|\Phi_j\rangle}\stat_{\mat{h}',\mat{\Lambda},\lambda}
\biggl\lbrace
\sum_jP_j\langle\Phi_j|\hat{W}|\Phi_j\rangle
\nonumber\\
&+&\frac{1}{\beta}\sum_j P_j\ln[P_j]
\nonumber\\
&-&\sum_{i,j}\Lambda_{i,j}\Bigl(\langle\Phi_j|\Phi_i\rangle-\delta_{j,i}\Bigr)
-\lambda\Bigl(\sum_j P_j-1\Bigr)
\nonumber\\
&+&\sum_{a,b}h'_{a,b}\Bigl(
\sum_j P_j\langle\Phi_j|\hat{c}^\dagger_a\hat{c}_b|\Phi_j\rangle-\rho_{b,a}
\Bigr)
\biggr\rbrace\;.
\label{eq:dmfphi}
\end{eqnarray}
The density-matrix functional $F_\beta^{\hat{W}}$ is the minimum of
the interaction energy and the entropy term. Note, that the
density-matrix functional does not vanish in the non-interacting limit
but that it contributes at finite temperature the entropy term
\begin{eqnarray}
F^{\hat{0}}_\beta[\mat{\rho}]=\frac{1}{\beta}\Tr\Bigl[
\mat{\rho}\ln(\mat{\rho})
-(\mat{1}-\mat{\rho})\ln(\mat{1}-\mat{\rho})\Bigr]
\end{eqnarray}
of the non-interacting electron gas.

An important property of the density-matrix functional is its
universality: The density-matrix functional is independent of the
one-particle Hamiltonian, because the latter, i.e. $\mat{h}'$ enters
only as a Lagrange multiplier.

As shown in Appendix~\ref{app:dmfconvexity}, the density-matrix
functional, defined as constrained search over ensembles, is convex.

The density matrix can be diagonalized in the one-particle Hilbert
space,
\begin{eqnarray}
\sum_{b}\rho_{a,b}\psi_{b,n}=\psi_{a,n}f_n
\;,
\end{eqnarray}
which produces the so-called natural
orbitals\cite{loewdin55_pr97_1474} $|\psi_n\rangle=\sum_a
|\chi_a\rangle\psi_{a,n}$ from the one-particle basisstates
$|\chi_a\rangle$ as eigenstates of the density matrix and the
occupations $f_n$ as its eigenvalues. The natural orbitals are
orthonormal one-particle states, such that
$\langle\psi_m|\psi_n\rangle=\delta_{m,n}$.

The density matrix must obey the so-called $N$-representability
property, that is it must be a matrix that can be obtained from an
ensemble of (fermionic) many-particle wave
functions. Coleman\cite{coleman63_rmp35_668} has shown that this
condition is identical to the requirement that the density matrix be
Hermitian and its eigenvalues, the occupations, lie between zero and
one.

The grand potential can be expressed more conveniently by the natural
orbitals and their occupations rather than directly by the
one-particle density matrix. In this representation, the grand
potential has a form that reminds us of density-functional theory,
\begin{eqnarray}
\Omega_{\beta,\mu}(\hat{h}+\hat{W})&=&
\min_{|\psi_n\rangle,f_n\in[0,1]}\stat_{\mat{\Lambda}}\biggl\lbrace
\sum_n f_n\langle\psi_n|\hat{h}|\psi_n\rangle
\nonumber\\
&+&F^{\hat{W}}_{\beta}\Bigl[\sum_n|\psi_n\rangle f_n\langle\psi_n|\Bigr]
-\mu\sum_n f_n
\nonumber\\
&-&\sum_{m,n}\Lambda_{m,n}\Bigl(\langle\psi_n|\psi_m\rangle-\delta_{n,m}\Bigr)
\biggr\rbrace
\;.
\label{eq:gcpdm}
\end{eqnarray}
Note, that we used the symbol $\mat{\Lambda}$ here in a different
context than before: Instead of constraining many-particle wave
functions to be orthonormal, here the orthonormality is enforced for
one-particle wave functions.  The main difference of
Eq.~\eqref{eq:gcpdm} from the corresponding expression for the
density-functional theory is that density-matrix functional theory
uses the true kinetic energy, while density-functional theory uses the
kinetic energy of a non-interacting electron gas.

\subsection{Kadanoff-Baym functional}
After having introduced the expression of the grand
potential using the density-matrix functional, let us now turn to the
alternative formulation in terms of Green's functions.

Luttinger and Ward\cite{luttinger60_pr118_1417} have shown that 
the grand potential can also be expressed as a functional of
the Matsubara Green's function and the self energy,
\begin{eqnarray}
\Omega_{\beta,\mu}(\hat{h}+\hat{W})=
\stat_{G,\Sigma}\Psi^{KB}_{\beta,\mu}[\mat{G},\mat{\Sigma},\mat{h},\hat{W}]
\;,
\label{eq:omegafrompsikb}
\end{eqnarray}
where $\Psi^{KB}_{\beta,\mu}$ is called the Kadanoff-Baym functional.
We adopt this naming following Chitra and Kotliar
\cite{chitra01_prb63_115110}.  The Green's function and the self
energy are given at fermionic Matsubara frequencies
$\omega_\nu=(2\nu+1)\pi/(\hbar\beta)$, specified by an integer $\nu$,
and obtained by a Fourier transformation
\begin{eqnarray}
G(i\omega_\nu)=\frac{1}{2}\int_{-\hbar\beta}^{\hbar\beta}d\tau\;
G(\tau)\e{i\omega_\nu\tau}
\end{eqnarray}
from of their imaginary-time partners. \cite{fetter71_book} The
imaginary-time Green's function, obeying the stationary condition, is
given by
\begin{eqnarray}
G_{\alpha,\beta}(\tau)=-\frac{1}{\hbar}\Tr\biggl\lbrace
\e{-\beta(\hat{h}+\hat{W}-\mu\hat{N}-\Omega)}
\mathcal{T}_\tau \Bigl\lbrace\hat{c}_\alpha(\tau)\hat{c}^\dagger_\beta(0)
\Bigr\rbrace
\biggr\rbrace
\;,
\nonumber \\
\label{eq:defgreensfunction}
\end{eqnarray}
where 
\begin{eqnarray}
\hat{c}_\alpha(\tau)&=&
\e{(\hat{h}+\hat{W}-\mu\hat{N})\tau/\hbar}
\hat{c}_\alpha
\e{-(\hat{h}+\hat{W}-\mu\hat{N})\tau/\hbar} 
\\
\hat{c}^\dagger_\alpha(\tau)&=&
\e{(\hat{h}+\hat{W}-\mu\hat{N})\tau/\hbar}
\hat{c}^\dagger_\alpha
\e{-(\hat{h}+\hat{W}-\mu\hat{N})\tau/\hbar} \; .
\end{eqnarray}
The self energy $\mat{\Sigma}(\tau)$ is defined\cite{fetter71_book}
such that it connects the two-particle Green's function to the
one-particle Green's function in the equation of motion for the
Green's functions by
\begin{eqnarray}
&&\sum_\gamma\int_0^{\hbar\beta} d\tau'\;
\Sigma_{\alpha,\gamma}(\tau-\tau')G_{\gamma,\beta}(\tau',0)
\nonumber\\
&=&\sum_{b,c,d} W_{\alpha,b,c,d}
\left(\frac{-1}{\hbar}\right)\Tr\biggl\lbrace
\e{-\beta(\hat{h}+\hat{W}-\mu\hat{N}-\Omega)}
\nonumber\\
&\times&\mathcal{T}_\tau\Bigl\lbrace
c^\dagger_b(\tau)\hat{c}_c(\tau)\hat{c}_d(\tau)\hat{c}_\beta^\dagger(0)\Bigr\rbrace
\biggr\rbrace
\;.
\label{eq:defselfenergy}
\end{eqnarray}

The Kadanoff-Baym functional\cite{luttinger60_pr118_1417}
\begin{eqnarray}
&&\hspace{-0.5cm}
\Psi^{KB}_{\beta,\mu}[\mat{G},\mat{\Sigma},\mat{h},\hat{W}] =
\Phi^{LW}_{\beta}[\mat{G},\hat{W}]
\nonumber\\&&
-\frac{1}{\beta}\sum_\nu 
\Tr\biggl\lbrace \ln\left(
1-\frac{1}
{(i\hbar\omega_\nu+\mu)\mat{1}-\mat{h}}
\mat{\Sigma}(i\omega_\nu)  \right)
\nonumber\\&&
+\mat{\Sigma}(i\omega_\nu) \mat{G}(i\omega_\nu) \biggr\rbrace
-\frac{1}{\beta}\Tr\Bigl\lbrace\ln\left(\mat{1}+\e{-\beta(\mat{h}-\mat{1}\mu)}\right)\Bigr\rbrace
\;.
\label{eq:kadanoffbaym2}
\end{eqnarray}
is built such that Green's function
Eq.~\eqref{eq:defgreensfunction} and self energy
Eq.~\eqref{eq:defselfenergy} result from the stationary conditions
specified in Eq.~\eqref{eq:omegafrompsikb}.

The first term on the right-hand side in Eq.\ \eqref{eq:kadanoffbaym2}
is the Luttinger-Ward functional $\Phi^{LW}_\beta$.  The
Luttinger-Ward functional is a sum of all closed,
connected and irreducible skeleton diagrams with the non-interacting
propagator replaced by the fully interacting
one.\cite{luttinger60_pr118_1417,abrikosov64_book} As is obvious
from its diagrammatic definition, it is a functional of the Green's
function and the interaction only but does not depend on the
one-particle Hamiltonian which contains the external potential.  In
this respect it is ``universal.''  Note that $\Phi^{LW}_\beta$
vanishes for a non-interacting system, that is for $\hat{W}=0$.
Furthermore, the diagrammatic construction implies
\begin{eqnarray}
\frac{\beta\delta\Phi^{LW}_\beta[\mat{G},\hat{W}]}{\delta{G}_{b,a}(i\omega_\nu)}
=\Sigma_{a,b}(i\omega_\nu) \: .
\label{eq:lwandselfenergy}
\end{eqnarray}
This equation is also equivalent with the stationary condition of the
Kadanoff-Baym functional with respect to the Green's function.  The
stationary condition with respect to the self energy is
\begin{eqnarray}
\mat{G}(i\omega_\nu)
=\Bigl(
(i\hbar\omega_\nu+\mu)\mat{1}-\mat{h}-\mat{\Sigma}(i\omega_\nu)
\Bigr)^{-1} \: .
\label{eq:dyson}
\end{eqnarray}
This is just Dyson's equation. 

\section{Connecting density-matrix functional and Luttinger-Ward functional}
\label{sec:kbtordmft}
Before we connect the Luttinger-Ward functional to the density-matrix
functional, we investigate the transformation properties of the
Kadanoff-Baym functional under changes $\mat{\Delta}$ of the
non-interacting Hamiltonian. 

In Appendix~\ref{app:invariance}, we show that
\begin{eqnarray}
\Psi^{KB}[\mat{G},\mat{\Sigma},\mat{h},\hat{W}]
&=&\Psi^{KB}[\mat{G},\mat{\Sigma}+\mat{\Delta},\mat{h}-\mat{\Delta},\hat{W}]
\nonumber\\
&&\hspace{-1cm}
+\frac{1}{\beta}\sum_\nu\e{i\beta\hbar\omega_\nu0^+}\Tr
\Bigl\lbrace\mat{G}(i\omega_\nu)\mat{\Delta}
\Bigr\rbrace\;.
\label{eq:transformationpsikb}
\end{eqnarray}
The important feature of Eq.~\eqref{eq:transformationpsikb} is that it
holds point-per-point and not only when the stationary conditions are
satisfied.  This allows one to choose the transformation depending on
the actual value of the Green's function $\mat{G}$.

In the following, we will employ this invariance, i.e.
Eq.~\eqref{eq:transformationpsikb}, to connect the Kadanoff-Baym
functional to the density-matrix functional by introducing a new
non-interacting Hamiltonian, which eventually will be a functional of
the density matrix.

A new non-interacting Hamiltonian
$\mathrho\left[\mat{\rho}[\mat{G}]\right]=\mat{h}-\mat{\Delta}\left[\mat{\rho}[\mat{G}]\right]$
  as functional of the Green's function is obtained as follows: From
  the Green's function $\mat{G}$ we obtain the one-particle reduced
  density matrix via
\begin{eqnarray}
\mat{\rho}[\mat{G}]=\frac{1}{\beta}\sum_\nu
\e{i\beta\hbar\omega_\nu0^+}
 \mat{G}(i\omega_\nu)
\;.
\label{eq:onepdenmatfromgreen}
\end{eqnarray}

The (non-interacting) Hamiltonian $\hathrho$ is defined by
requiring that it lead to the density matrix given by
Eq.~\eqref{eq:onepdenmatfromgreen}, i.e.:
\begin{eqnarray}
\mat{\rho}=\biggl[\mat{1}+\e{\beta(\mathrho-\mu\mat{1})}\Bigr]^{-1}
\;,
\label{eq:fermi}
\end{eqnarray}
Explicitly, it is given by
\begin{eqnarray}
\mathrho[\mat{\rho}]=\mu\mat{1}
+\frac{1}{\beta}
\ln\Bigl[\frac{\mat{1}-\mat{\rho}}{\mat{\rho}}\Bigr]
\;.
\label{eq:defhrho}
\end{eqnarray}
The identity is easily seen in the representation of natural orbitals,
where all matrices are diagonal: Then, Eq.~\eqref{eq:fermi} is the
expression for the Fermi distribution. The Hamiltonian
$\mathrho$ is, in general, a non-local Hamiltonian.

The construction just described is practical only at finite
temperatures. At zero temperature the spectrum of $\mathrho$ collapses
to a single energy, namely the Fermi level.

The required change for the one-particle Hamiltonian is
\begin{eqnarray}
\mat{\Delta}[\mat{\rho}]=\mat{h}-\mathrho=
\Bigl(\mat{h}-\mu\mat{1}\Bigr)-\frac{1}{\beta}
\ln\Bigl[\frac{\mat{1}-\mat{\rho}}{\mat{\rho}}\Bigr]
\;.
\end{eqnarray}


The transformation turns the grand potential of non-interacting
electrons (last term in Eq.~\eqref{eq:kadanoffbaym2}) into a form that
can be expressed conveniently by the natural orbitals $|\psi_n\rangle$
and their occupations $f_n$, i.e. the eigenvectors and eigenvalues of
the reduced density matrix $\mat{\rho}$.
\begin{eqnarray}
&&\hspace{-1cm}-\frac{1}{\beta}\Tr\Bigl\lbrace\ln
\left(1+\e{-\beta(\mathrho-\mat{1}\mu)}\right)
\Bigr\rbrace
\nonumber\\
&=&
\sum_n f_n\langle\psi_n|\hathrho|\psi_n\rangle
-\mu\sum_n f_n
\nonumber\\
&+&\frac{1}{\beta} \sum_n \Bigl[ f_n \ln(f_n)+(1-f_n)\ln(1-f_n)\Bigr]
\;,
\label{eq:intronatorb}
\end{eqnarray}

With this, the Kadanoff-Baym functional can be rewritten as
\begin{widetext}
\begin{eqnarray}
\Psi^{KB}_{\beta,\mu}[\mat{G},\mat{\Sigma},\mat{h},\hat{W}]
&=&
\sum_n f_n\langle\psi_n|\hat{h}|\psi_n\rangle
+\frac{1}{\beta}\sum_n
\Bigl(f_n\ln(f_n)+(1-f_n)\ln(1-f_n)\Bigr)-\mu\sum_n f_n
\nonumber\\
&+&
\Phi^{LW}_\beta[\mat{G},\hat{W}]
-\frac{1}{\beta}\sum_\nu\e{i\beta\hbar\omega_\nu0^+}\Tr\Bigl\lbrace
\ln\Bigl[
\mat{1}-
\Bigl((i\hbar\omega_\nu+\mu)\mat{1}-\mathrho\Bigr)^{-1}
\Bigl(\mat{h}+\mat{\Sigma}(i\omega_\nu)-\mathrho\Bigr)
\Bigr]
\nonumber\\
&+&\Bigl(\mat{h}+\mat{\Sigma}(i\omega_\nu)-\mathrho\Bigr)\mat{G}(i\omega_\nu)
+
\mat{G}(i\omega_\nu)(\mathrho-\mat{h})\Bigr\rbrace
+\sum_n f_n\langle\psi_n|(\hathrho-\hat{h})|\psi_n\rangle
\;,
\label{eq:kbwithhrho}
\end{eqnarray}
\end{widetext}
where $\mathrho$, the natural orbitals $|\psi_n\rangle$ and the
occupations $f_n$ are, via Eq.~\eqref{eq:onepdenmatfromgreen},
functionals of the Green's function $\mat{G}$.  The last term in
Eq.~\eqref{eq:kbwithhrho} ensures that the first term is the expectation
value of the true one-particle Hamiltonian and not the expectation
value of $\hathrho$. The operator $\hathrho$ is defined via
the matrix $\mathrho$ analogously to Eq.~\eqref{eq:hath}. 

Note that the last term $\mat{G}(i\omega_\nu)(\mathrho-\mat{h})$
in the Matsubara sum in Eq.~\eqref{eq:kbwithhrho} can in principle be
canceled against a similar contribution in the preceding term.  It
however also combines with the last term of Eq.~\eqref{eq:kbwithhrho} to
a contribution that vanishes when the Green's function fulfills the
density-matrix constraint. By keeping it explicitly we maintain the
integrity of conceptually related entities and make the derivation
more transparent.

For practical purposes, it will be convenient to rewrite the last two
terms in Eq.~\eqref{eq:kbwithhrho} using the identity
\begin{eqnarray}
&-&\Tr\Bigl\lbrace
\Bigl(\mat{\rho}
-\frac{1}{\beta}\sum_\nu
 \e{i\beta\hbar\omega_\nu0^+}
 \mat{G}(i\omega_\nu)
 \Bigr)
\Bigl(\mat{h}-\mathrho\Bigr)\Bigr\rbrace
\nonumber\\
&=&
\frac{1}{\beta}\sum_\nu
\Tr\Bigl\lbrace
\left(
\mat{G}(i\omega_\nu)
-\frac{1}{(i\hbar\omega_\nu+\mu)\mat{1}-\mathrho}
\right)
\nonumber\\
&&
\times\Bigl(\mat{h}-\mathrho\Bigr)
\Bigr\rbrace\;,
\end{eqnarray}
because the latter has a converging Matsubara sum.

In the spirit of density-functional theory, we can treat the density
matrix, that is the natural orbitals and the occupations, as
independent variables that are, however, linked to the Green's function
by a constraint that enforces Eq.~\eqref{eq:onepdenmatfromgreen}.

This argument can also be turned around: Instead of constraining the
density matrix to the Green's function during the optimization of the
latter, we may as well perform a search over the density matrices. For
each density matrix, the optimum Green's function is determined under
a density-matrix constraint. This approach naturally leads to a
density-matrix functional expressed as a constrained search over
Green's functions and self-energies. The search over Green's functions
is guided by a stationarity principle rather than by an extremum principle.

From Eq.~\eqref{eq:kbwithhrho}, we obtain the grand potential
in the form
\begin{widetext}
\begin{eqnarray}
\Omega^{KB}_{\beta,\mu}[\hat{h}+\hat{W}]
&=&\min_{|\psi_n\rangle,f_n\in[0,1]}\stat_\Lambda
\biggl\lbrace\sum_n f_n\langle\psi_n|\hat{h}|\psi_n\rangle
+\tilde{F}_\beta^{\hat{W}}\Bigl[\sum_n|\psi_n\rangle f_n\langle\psi_n|\Bigr]
-\mu\sum_n f_n
-\sum_{m,n}\Lambda_{m,n}\Bigl(\langle\psi_n|\psi_m\rangle-\delta_{m,n}\Bigr)
\biggr\rbrace
\nonumber \\
\label{eq:gcpdm2}
\end{eqnarray}
with 
\begin{eqnarray}
\tilde{F}^{\hat{W}}_\beta[\mat{\rho}]
&=&
\frac{1}{\beta}\Tr\Bigl[
\mat{\rho}\ln(\mat{\rho})+(\mat{1}-\mat{\rho})\ln(\mat{1}-\mat{\rho})\Bigr]
\nonumber\\
&+&
\stat_{\mat{h}'}\stat_{\mat{G},\mat{\Sigma}}
\biggl\lbrace
\Phi^{LW}_\beta[\mat{G},\hat{W}]
-\frac{1}{\beta}\sum_\nu
\Tr\Bigl\lbrace
\ln\Bigl[
\mat{1}-
\Bigl((i\hbar\omega_\nu+\mu)\mat{1}-\mathrho\Bigr)^{-1}
\Bigl(\mat{h}'+\mat{\Sigma}(i\omega_\nu)-\mathrho\Bigr)
\Bigr]
\nonumber\\&&
+\Bigl(\mat{h}'+\mat{\Sigma}(i\omega_\nu)-\mathrho\Bigr)
\mat{G}(i\omega_\nu)
-
\Bigl[
\mat{G}(i\omega_\nu)
-\Bigl((i\hbar\omega_\nu+\mu)\mat{1}-\mathrho\Bigr)^{-1}
\Bigr]
\Bigl(\mat{h}'-\mathrho\Bigr)\Bigr\rbrace
\biggr\rbrace
\;.
\label{eq:dmfgreen}
\end{eqnarray}
\end{widetext}
where $\mathrho$ is a functional of the one-particle
reduced density matrix $\mat{\rho}$.

Let us explore the various terms in Eq.~\eqref{eq:dmfgreen}: The last
term in the Matsubara sum of Eq.~\eqref{eq:dmfgreen},
$\frac{1}{\beta}\sum_\nu(\mat{G}-\matGrho)(\mat{h}'-\mathrho)$
vanishes, when the Green's function $\mat{G}$ obeys the density-matrix
constraint. $\matGrho$ is a short hand for
\begin{eqnarray}
\matGrho(i\omega_\nu)
&=&\Bigl((i\hbar\omega_\nu+\mu)\mat{1}-\mathrho\Bigr)^{-1}
\;.
\label{eq:defgrho}
\end{eqnarray}

The Luttinger-Ward functional combined with the remainder
of the Matsubara sum corresponds the interaction part of the
Kadanoff-Baym functional for an interacting system with a one-particle
Hamiltonian $\mathrho$ instead of $\mat{h}$ and a self energy
$\mat{h}'+\mat{\Sigma}-\mathrho$.

In the derivation of Eq.~\eqref{eq:dmfgreen}, we exploited that the
derivative of the functional inside the expression of
Eq.~\eqref{eq:dmfgreen} with respect to the one-particle Hamiltonian
$\mat{h}'$ (formerly $\mat{h}$) vanishes when the density-matrix
constraint Eq.~\eqref{eq:onepdenmatfromgreen} is obeyed. Thus, the
constraint that the Green's function is consistent with the density
matrix can be imposed simply by requiring that the derivative with
respect to $\mat{h}'$ vanishes.  We have added a prime to make it
evident that $\mat{h}'$ is a Lagrange multiplier.

Like the density-matrix functional $F^{\hat{W}}_\beta$ defined via a
constrained search over many-particle wave functions, the functional
$\tilde{F}_\beta^{\hat{W}}$ is universal, that is, it is independent
of the one-particle Hamiltonian $\mat{h}$ of the physical system
  of interest and also independent of the chemical potential.

The fact that it is independent of the chemical
potential is seen from the definition of $\mathrho$ and from the
fact that 
$\mat{h'}$ 
is a Lagrange multiplier, which can absorb a
constant term. Thus, we may simply set the chemical potential in
Eq.~\eqref{eq:dmfgreen} and in Eq.~\eqref{eq:defhrho} to zero.

With Eq.~\eqref{eq:dmfgreen} and Eq.~\eqref{eq:dmfphi} we have two
very different representations of the density-matrix functional.  It
is most important to realize, however, that both representations are
exact representations of the same density-matrix functional.  This
becomes obvious from the comparison of Eq.~\eqref{eq:gcpdm2} with
Eq.~\eqref{eq:gcpdm}, from the fact that both representations are
universal and from the fact that, up to now, all calculations have
been free of any approximation.  The representation
Eq.~\eqref{eq:dmfgreen} is novel and makes the link between
Green's-function-based approaches and those based on many-particle wave
functions.

Having derived a density-matrix functional from Green's functions and
self-energies, we have introduced a true minimum principle into the
Green's function world. The search for the physical one-particle density
matrix has a true minimum principle. The determination of the minimum
is substantially simpler than the search for a saddle point.

Some caution is still required. The minimum property follows from the
representation Eq.~\eqref{eq:dmfphi} for the density-matrix functional
but is obviously guaranteed for the exact functional only, rather than
for every conceivable approximation. We consider this as a
mathematical caveat, that will not be relevant in practice:
Approximations that change the total energy surface qualitatively can
probably not be considered as reliable.

One should also note that it is unavoidable that one part of the
optimization still requires a saddle point search. This part has been
moved into the evaluation of the density-matrix functional from
Green's functions and self energies.

For the sake of completeness, let us note here the stationary
conditions of Eq.~\eqref{eq:dmfgreen} 
\begin{eqnarray}
\mat{\Sigma}(i\omega_\nu)&=&
\beta\frac{\delta\Phi^{LW}_\beta[\mat{G},\hat{W}]}{\delta\mat{G}(i\omega_\nu)}
\label{eq:stationaryG}
\\
\mat{\rho}&=&\frac{1}{\beta}\sum_\nu\e{i\beta\hbar\omega_\nu0^+}
\mat{G}(i\omega_\nu)
\label{eq:stationaryhprime}
\\
\mat{G}(i\omega_\nu)
&=&\Bigl[(i\hbar\omega_\nu+\mu)\mat{1}
-\mat{h}'-\mat{\Sigma}(i\omega_\nu)\Bigr]^{-1}
\;.
\label{eq:stationarysigma}
\end{eqnarray}
Furthermore, as shown in Appendix~\ref{app:dfdrho}, the functional
derivative of the density-matrix functional Eq.~\eqref{eq:dmfgreen} is
\begin{eqnarray}
\frac{\partial\tilde{F}_\beta^{\hat{W}}[\mat{\rho}]}{\partial\mat{\rho}}
&=&-(\mat{h}'-\mu)
\;.
\label{eq:derivativedensitymatrixfunctional}
\end{eqnarray}

Combining this with the stationary condition of the Kadanoff-Baym
functional Eq.~\eqref{eq:gcpdm2} with respect to variations of the
density matrix we obtain
\begin{eqnarray}
\mat{h}=\mat{h}'
\label{eq:minimumrdmf}
\end{eqnarray}
as condition for the minimum.

The expression in the density-matrix functional
$\tilde{F}_\beta^{\hat{W}}$ is not an extremum with respect to the
self energy or the Green's function.  Therefore, we cannot use a
gradient-following technique, but we have to resort to a
self-consistency scheme defined by the following steps.

\begin{enumerate}
\item First one constructs the Hamiltonian $\mathrho$ from the
  specified density matrix $\mat{\rho}$ via Eq.~\eqref{eq:defhrho}. The
  value for the chemical potential is arbitrary and can be set to
  zero.
\item The initial self energy is set to zero and the
  Lagrange multiplier $\mat{h}'$ is set equal to
  $\mathrho$. This ensures that the initial Green's function
  \begin{eqnarray}
      \mat{G}(i\omega_\nu)=\Bigl[(i\hbar\omega_\nu+\mu)\mat{1}
-\mat{h}'-\mat{\Sigma}(i\omega_\nu)\Bigr]^{{-1}}
   \label{eq:gofhprime}
  \end{eqnarray}
  is equal to $\matGrho$ defined in Eq.~\eqref{eq:defgrho},
  which satisfies the density-matrix constraint.
\item Next we evaluate a new self-energy from
  Eq.~\eqref{eq:lwandselfenergy}. 
\item The new self energy and $\mat{h}'$ define a Green's function
  $\bar{\mat{G}}(i\omega_\nu)$ via Dyson's equation
  Eq.~\eqref{eq:gofhprime}, which does not yet satisfy the
  density-matrix constraint. Now the Lagrange multiplier $\mat{h}'$ is
  adjusted such that the constraint condition is fulfilled.  This is
  done iteratively. In each iteration, the constraint equation is
  expanded to first order in the change $\delta\mat{h}'$ of the
  Lagrange multiplier $\mat{h}'$.  This linearized constraint
  condition
  \begin{eqnarray}
   \frac{1}{\beta}\sum_{\nu}\e{i\beta\hbar\omega0^+}
   \Bigl(\bar{\mat{G}}(i\omega_\nu)
   +\bar{\mat{G}}(i\omega_\nu)\delta\mat{h}'\bar{\mat{G}}(i\omega_\nu)
\Bigr)=\mat{\rho}
\nonumber\\
  \end{eqnarray}
  provides a correction $\delta\mat{h}'$ to the Lagrange multiplier.
  The new Lagrange multiplier defines an improved Green's function
  $\bar{\mat{G}}(i\omega_\nu)$ via Eq.~\eqref{eq:gofhprime}. This step 4
  is repeated until the Lagrange multiplier is converged.
\item Once the constraint is obeyed, $\mat{G}$ is replaced by the
  converged $\bar{\mat{G}}$ and one returns to step 3 and recalculates
  the self energy.
\item When the loop of steps 3-5 is converged, the energy contribution
  from the density-matrix functional is evaluated from
  Eq.~\eqref{eq:dmfgreen} and its derivative from
  Eq.~\eqref{eq:derivativedensitymatrixfunctional}.
\end{enumerate}
%
\section{Approximations}
\label{sec:approximations}
One of the major benefits of the new construction of the
density-matrix functional defined in Eq.~\eqref{eq:dmfgreen}
is that it not only makes contact with density-functional theory but also
with approximations defined diagrammatically in the context of 
Green's-function approaches.

\subsection{Relation to density-functional theory}
There is an intimate connection between density-matrix functional
theory and density-functional theory: It is easily shown that the
density functional can be obtained from a constrained minimization of
the density-matrix functional.  However, in order to embed an explicit
treatment of many-particle correlations into the available density
functional codes as in methods that link DFT with dynamical mean-field
theory \cite{lechermann06_prb74_125120,aichhorn09_prb80_85101,
  haule10_prb81_195107,amadon12_jpcm24_75604}, it is desirable to take
the opposite route, namely to develop an approximate density-matrix
functional that is consistent with a given density functional.

The grand potential in density-functional theory
is\cite{mermin65_pr137_A1441}
\begin{eqnarray}
\Omega_{\beta,\mu}[v_{ext}]&=&\min_{n}\biggl\lbrace
K_s[n]+\int d^3r\;v_{ext}(\vec{r})n(\vec{r})
\nonumber\\
&&\hspace{-1cm}
+E_H[n]
+E_{xc,\beta}[n] -\mu \int d^3r\;n(\vec{r})\biggr\rbrace
\label{eq:dftetot}
\end{eqnarray}
where $K_s[n]$ is the intrinsic energy, i.e., kinetic energy and
entropy term, of a non-interacting electron gas with density
$n(\vec{r})$ at a temperature specified by $\beta$.  The intrinsic
energy of the interacting electrons contains in addition the Hartree
energy $E_H[n]$ and the exchange correlation energy $E_{xc}[n]$.  In
Appendix~\ref{app:adiabatic}, we briefly review the adiabatic
connection\cite{harris84_pra29_1648} to make the notation more
explicit, with a particular emphasis on the extension to finite
temperatures.  The exchange-correlation energy $E_{xc,\beta}[n]$
contains at finite temperatures not only a contribution of the kinetic
energy but also an additional correction to account for the entropy
difference of the interacting and the non-interacting electron gas.

Comparing Eq.~\eqref{eq:dftetot} with
Eq.~\eqref{eq:rdmftgrandcanonicalpotouter} of the density-matrix
functional theory, we can construct a new density-matrix functional.
\begin{eqnarray}
\tilde{F}^{\hat{W},DFT}_{\beta}[\mat{\rho}]
&=&
E_{H}[n]+E_{xc,\beta}[n]+K_s[n]
-E_{kin}[\mat{\rho}]
\;,
\nonumber\\
\label{eq:dmffromdft}
\end{eqnarray}
where $n(\vec{r})$ is obtained from $\mat{\rho}$.  

Even for the exact density functional, the functional
$\tilde{F}^{\hat{W},DFT}_{\beta}$ constructed in this way differs from
the exact density-matrix functional $F^{\hat{W}}_{\beta}$ defined
earlier in Eq.~\eqref{eq:dmfphi}.  
The two functionals agree, however,
for the physically relevant density matrices $\mat{\rho}[n]$, which
are obtained from minimization of $F^{\hat{W}}_{\beta}[\mat{\rho}]$
for a fixed density $n(\vec{r})$, i.e.
\begin{eqnarray}
E_{kin}[\mat{\rho}]+F^{\hat{W},DFT}_\beta[\mat{\rho}]&=&
\min_{\mat{\rho'}\rightarrow n[\mat{\rho}]}
E_{kin}[\mat{\rho'}]+F^{\hat{W}}_\beta[\mat{\rho'}]\;.
\nonumber\\
\end{eqnarray}
Thus, both functionals predict the
same grand potentials.


Therefore, we consider this density-matrix
functional derived from density-functional theory as a useful
expression to obtain double-counting correction terms for the
embedding of higher-level theories into density-functional theory with
conventional, parametrized density functionals.

\subsection{Hartree-Fock Approximation}
The Hartree-Fock approach can be considered as resulting from the Ritz
variational principle using a Slater determinant as a trial many-body
wave function.  Alternatively, the approach can be defined by the most
simple truncation of the Luttinger-Ward functional.  Keeping only
first-order renormalized skeleton diagrams, we find 
\begin{eqnarray}
  \Phi^{LW,HF}[\mat{G}]
  &=&
  \frac{1}{2}
  \frac{1}{\beta^{2}} 
  \sum_{\nu,\nu'} e^{i \beta \hbar \omega_\nu 0^{+}} e^{i \beta \hbar \omega_{\nu'} 0^{+}}
  \sum_{a,b,c,d}
  U_{a,b,d,c} 
  \nonumber\\ 
  &&\hspace{-1.2cm}\times\Big(
  G_{d,a}(i\omega_\nu)G_{c,b}(i\omega_{\nu'})
  -  G_{c,a}(i\omega_\nu)G_{d,b}(i\omega_{\nu'})
  \Big) \: .
  \nonumber\\ 
\end{eqnarray}
Carrying out the summations over the Matsubara frequencies, this
trivially becomes a density-matrix functional.  

After satisfying the stationary conditions Eqs.~\eqref{eq:stationaryG},
\eqref{eq:stationaryhprime} and ~\eqref{eq:stationarysigma}, the
density-matrix functional Eq.~\eqref{eq:dmfgreen} in the Hartree-Fock
approximation is
\begin{eqnarray}
\tilde{F}^{\hat{W},HF}_\beta[\mat{\rho}]
&=&
\frac{1}{2}\sum_{a,b,c,d}U_{a,b,d,c}
\Bigl(\rho_{d,a}\rho_{c,b}-\rho_{c,a}\rho_{d,b}\Bigr)
\nonumber\\
&+&
\frac{1}{\beta}\Tr\Bigl[
\mat{\rho}\ln(\mat{\rho})+(\mat{1}-\mat{\rho})\ln(\mat{1}-\mat{\rho})\Bigr]
\;,
\label{eq:rdmfhf}
\end{eqnarray}
which consists of Hartree and exchange energy as well as an entropy
contribution.

The Hartree-Fock self energy, obtained from
Eq.~\eqref{eq:stationaryG} is frequency independent and equals
$\Sigma^{HF}_{a,b}[\mat{\rho}]$, which has the form
\begin{eqnarray}
  \Sigma^{HF}_{a,b}[\mat{\rho}]=
  \sum_{c,d}
  ( U_{a,c,b,d} - U_{a,c,d,b} ) \rho_{d,c}\;.
\label{eq:hfselfenergy}
\end{eqnarray}

The optimum Green's function, which satisfies
Eq.~\eqref{eq:stationarysigma}, is
\begin{eqnarray}
\mat{G}^{HF}(i\omega_\nu)=
\Bigl[(i\hbar\omega_\nu+\mu)
\mat{1}-\mat{h}'-\mat{\Sigma}^{HF}[\mat{\rho}]\Bigr]^{-1}
\;,
\label{eq:hfgreensfunction}
\end{eqnarray}
where the Lagrange multiplier $\mat{h}'$ is chosen
such that the density-matrix constraint
Eq.~\eqref{eq:stationaryhprime} is obeyed.

The Lagrange multiplier can also obtained from
Eq.~\eqref{eq:derivativedensitymatrixfunctional} by inserting the
derivative of the density-matrix functional Eq.~\eqref{eq:rdmfhf} with
respect to the density matrix
\begin{eqnarray}
\mat{h'}
&=&\mu\mat{1}-\mat{\Sigma}^{HF}[\mat{\rho}]
-k_BT\ln\frac{\mat{\rho}}{\mat{1}-\mat{\rho}}\;.
\end{eqnarray}
This identifies the Hartree-Fock Green's function
Eq.~\eqref{eq:hfgreensfunction} with $\matGrho$ defined earlier in
Eq.~\eqref{eq:defgrho}.

For a given one-particle Hamiltonian, the minimum condition
Eq.~\eqref{eq:minimumrdmf} of the grand potential with respect
to the density matrix provides us with the optimum density matrix
\begin{eqnarray}
\mat{\rho}=
\Bigl(\mat{1}+\e{\beta(\mat{h}+\mat{\Sigma}^{HF}[\mat{\rho}]-\mu\mat{1})}
\Bigr)^{-1}\;,
\end{eqnarray}
which must be solved self-consistently with Eq.~\eqref{eq:hfselfenergy}.

The Hartree-Fock approximation is closely related to the
  LDA+$U$ method\cite{anisimov91_prb44_943,liechtenstein95_prb52_5467}
  and the hybrid density
  functionals\cite{becke93_jcp98_1372,becke93_jcp98_5648}, which
  replace part of the local exchange of DFT by the exact Fock
  term. Both methods can be considered as hybrid methods of
  density-functional theory and the Hartree-Fock approximation in the
  sense described earlier.\cite{bloechl11_prb84_205101}

The Hartree-Fock approximation is the most simple example of a
so-called ``conserving approximation'' in the sense of Baym and
Kadanoff. \cite{baym61_pr124_287,baym62_pr127_1391} Conserving
approximations result from truncations of the Luttinger-Ward
functional and respect the macroscopic conservation laws resulting
from continuous symmetries of the Hamiltonian.  There have been a
variety of attempts to systematically improve on the Hartree-Fock
approximation by taking into account additional classes of diagrams
from self-consistent second-order perturbation theory, the random-phase
approximation and the self-consistent $GW$ approach,\cite{hedin65_pr139_796}
over the T-matrix or fluctuation-exchange
approximation\cite{baym61_pr124_287,bickers89_ap193_206} to rather
complex theories such as the Parquet
equations.\cite{dedominicis64_jmathphys5_14,bickers89_ap193_206} In
the present context, they provide approximate, namely perturbatively
defined, density-matrix functionals which should be valid for ``weakly
correlated'' systems.

\subsection{Density-matrix functionals without Green's function 
optimization}
\label{sec:nde2}
One of the main computational obstacles related to density functionals
derived from the Kadanoff-Baym functional is that they are defined
through a set of self-consistency equations involving Green's
functions and self energies. An algorithmic expression for the
density-matrix functional that can be evaluated directly, as that of
the Hartree-Fock calculation, would thus be highly desirable.

Here, we present such an approximation, which results from the neglect
of dynamical effects to second and higher orders (NDE2). To obtain
this approximation, we rewrite the density-matrix functional
Eq.~\eqref{eq:dmfgreen} as
\begin{eqnarray}
\tilde{F}^{\hat{W}}_\beta[\mat{\rho}]
&=&
\frac{1}{\beta}\Tr\Bigl[
\mat{\rho}\ln(\mat{\rho})+(\mat{1}-\mat{\rho})\ln(\mat{1}-\mat{\rho})\Bigr]
\nonumber\\
&+&\Phi^{LW}_\beta[\matGrho[\mat{\rho}],\hat{W}]
+\Delta\tilde{F}_\beta^{\hat{W}}[\mat{\rho}]
\;,
\end{eqnarray}
where the argument $\matGrho[\mat{\rho}]$ of the Luttinger-Ward
functional is a direct functional of the density matrix defined in
Eq.~\eqref{eq:defgrho}. The remainder
\begin{eqnarray}
\Delta\tilde{F}_\beta^{\hat{W}}[\mat{\rho}]&=&
\Phi^{LW}_\beta[\mat{G},\hat{W}]
-\Phi^{LW}_\beta[\matGrho,\hat{W}]
\nonumber\\
&+&\sum_\nu\Tr\Bigl\lbrace
\left.\frac{\partial\Phi^{LW}_\beta}{\partial\mat{G}(i\omega_\nu)}
\right|_{\mat{G}}
\Bigl(\matGrho(i\omega_\nu)-\mat{G}(i\omega_\nu)\Bigr)
\Bigr\rbrace
\nonumber\\
&+&\frac{1}{\beta}\sum_\nu\Tr\Bigl\lbrace
\sum_{n=2}^\infty\frac{1}{n}
\Bigl(
\mat{1}-\matGrho(i\omega_\nu)
\mat{G}^{-1}(i\omega_\nu)\Bigr)^n
\Bigr\rbrace\;,
\nonumber\\
\end{eqnarray}
contains only terms that are quadratic or of higher order in
$\mat{G}[\mat{\rho}]-\matGrho[\mat{\rho}]$, where
$\mat{G}=\mat{G}[\mat{\rho}]$ is the Green's function that obeys the
stationary conditions for the specified density matrix $\mat{\rho}$.

If the dynamic correlations embodied in the deviation
$\mat{G}[\mat{\rho}]-\matGrho[\mat{\rho}]$ are small, we can ignore
the remainder $\Delta\tilde{F}_\beta^{\hat{W}}[\mat{\rho}]$.  This
constitutes the NDE2 approximation, which has the advantage that
$\matGrho$ is obtained directly from the density matrix. Thus, the
algorithmic complexity of NDE2 is similar to that of the Hartree-Fock
approximation, in the sense that self-consistency loops are
  avoided and that only Green's functions with static self energies
  need to be considered.  In contrast to the Hartree Fock
approximation, however, it allows to systematically include
higher-order terms in the interaction through the choice of the
Luttinger-Ward functional.  The NDE2 approximation is compatible with
any approximation of the Luttinger-Ward functional.

The NDE2 approximation only affects the difference between two Green's
function with the same density matrix.  In contrast to expansions in
the interaction, the terms, which are ignored in the NDE2
approximation do not affect the kinetic energy, the electron density
nor the exchange hole. The approximations of NDE2 are from the outset
limited strictly to the shape of the correlation hole. Thus, the NDE2
approximation may profit from a preservation of these sum rules.

Furthermore, the Green's function entering the Luttinger-Ward
functional is a Green's function for a non local but static
potential.  This simplifies the calculations to some extent.

Thus, the apparent advantage of the NDE2 approximation is that it
allows us to systematically construct approximations of the density
matrix functional Eq.~\eqref{eq:dmfphi} from established many-body
theories, keeping the complexity within a reasonable margin. To the
best of our knowledge, the NDE2 is the first such approximation.

\subsection{Dynamical mean-field theory}
\label{sec:dynamicalmeanfieldtheory}
Dynamical mean-field theory can be defined as an approximation to the
Luttinger-Ward functional and represents a \textit{non-perturbative}
conserving approximation.\cite{georges96_rmp68_13}

In dynamical mean-field theory, one usually limits the interaction to
site-local terms. That is, clusters $\mathcal{C}_R$ of local orbitals
are defined that we name \textit{correlated Hilbert spaces}, and the
interaction tensor $U_{a,b,c,d}$ is limited to these local Hilbert
spaces.  This amounts to the approximation
\begin{eqnarray}
\hat{W}\approx\sum_R \hat{W}_R
\label{eq:localapprox}
\end{eqnarray}
with
\begin{eqnarray}
\hat{W}_R=\frac{1}{2}\sum_{a,b,c,d\in\mathcal{C}_R}
U_{a,b,d,c}\hat{c}_a^\dagger\hat{c}_b^\dagger
\hat{c}_c\hat{c}_d \: .
\end{eqnarray}
The resulting Hamiltonian is that of a multi-band Hubbard model.  In
order to compensate for this truncation, the U-tensor elements are
usually scaled down. 

The Luttinger-Ward functional is only sensitive to the
Green's-function matrix elements that are directly connected to the
U-tensor elements. Thus, the self energy, which is given as the
derivative of the Luttinger-Ward functional by
Eq.~\eqref{eq:lwandselfenergy}, only acts on the union of the correlated
Hilbert spaces.

The defining approximation of dynamical mean-field theory, ontop of
the restriction to Hubbard-like Hamiltonians, is the local
approximation of the self energy, which divides the self energy into a
sum of local terms. Formulated in terms of the Luttinger-Ward
functional, this means that the corresponding Luttinger-Ward
functional is a sum of terms
\begin{eqnarray}
\Phi^{LW}_\beta[\mat{G},\sum_R\hat{W}_R]
&\approx&
\sum_R
\Phi^{LW}_\beta[\mat{G},\hat{W}_R] \: .
\label{eq:approxdmft}
\end{eqnarray}
Because only Green's function elements contribute to the
Luttinger-Ward functional that are connected to an interaction, this
approximation implies that each term
$\Phi^{LW}_\beta[\mat{G},\hat{W}_R]$ is sensitive only to those
elements of the Green's function that connect orbitals within the same
local cluster $\mathcal{C}_R$.

In this approximation, the self energy 
\begin{eqnarray}
\mat{\Sigma}(i\omega_\nu)=\sum_{R}
\beta
\frac{\delta\Phi^{LW}_\beta[\mat{G},\hat{W}_R]}{\delta\mat{G}(i\omega_\nu)} \: .
\label{eq:dmftlocalself1}
\end{eqnarray}
is a sum over the correlated Hilbert spaces. 

While each term in the sum of Eq.~\eqref{eq:dmftlocalself1} is formally
a matrix in the full one-particle Hilbert space,
the self energy can also be written as the direct sum 
\begin{eqnarray}
\mat{\Sigma}(i\omega_\nu)=\bigoplus_{R}\mat{\Sigma}_R(i\omega_\nu)
\end{eqnarray}
of local self energies 
\begin{eqnarray}
\mat{\Sigma}_R(i\omega_\nu)=\beta
\frac{\delta\Phi^{LW}_\beta[\mat{G}_R,\hat{W}_R]}
{\delta\mat{G}_R(i\omega_\nu)} \: .
\label{eq:dmftlocalself}
\end{eqnarray}
where each local self energy $\mat{\Sigma}_R$ or Green's function
$\mat{G}_R$ is a matrix with finite dimension in the specific
correlated Hilbert space $\mathcal{C}_R$. On the correlated Hilbert
space $\mathcal{C}_R$, the local Green's function $\mat{G}_R$ is
identical to $\mat{G}$ and the local self energy $\mat{\Sigma}_R$ is
identical to $\mat{\Sigma}$.

As a result of the local approximation of the self energy, the latter
has no dependence on the reciprocal wave vector in a representation of
Bloch waves.

When dynamical mean-field theory is formulated as limiting theory for
infinite coordination number\cite{metzner89_prl62_324},
M\"uller-Hartmann has shown for a class of Hamiltonians with non-local
electrostatic interaction that the Hartree energy remains finite,
while the non-local exchange-correlation energy
vanishes\cite{muellerhartmann89_zpb74_507}. Because the Hartree
energy, like the exchange energy, is a direct functional of the
density matrix, these terms are easily separated out in a
density-matrix functional framework, so that they do not add to the
complexity of the formalism.

\section{Forces and spectral properties}
\label{sec:forces}
%
\subsection{Forces}
The force $F_i$ on the atomic coordinate $R_i$ is given by the
derivative of the Helmholtz free energy
$A_{\beta,N}:=\Omega_{\beta,\mu}+\mu N$.  The Helmholtz free energy is
obtained from Eq.~\eqref{eq:gcpdm2} as
\begin{eqnarray}
A_{\beta,N}(\vec{R})
&=&
\Tr\Bigl[\mat{\rho}(\vec{R})\mat{h}(\vec{R})\Bigr]
+\tilde{F}_\beta^{\hat{W}(\vec{R})}\Bigl[\mat{\rho}(\vec{R})\Bigr]\;,
\end{eqnarray}
where $\mat{\rho}(\vec{R})$ fulfills the minimum and stationary
conditions.  The atomic positions enter via the Hamiltonian matrix
elements, i.e. via $\mat{h}$ and $U_{a,b,c,d}$.  The corresponding
force is therefore
\begin{eqnarray}
F_i&=&
-\Tr[\frac{\partial\mat{h}}{\partial R_i}\mat{\rho}]
-
\sum_{a,b,c,d}
\frac{\partial\Phi^{LW}_\beta}{\partial U_{a,b,c,d}}
\frac{\partial U_{a,b,c,d}}{\partial R_i}
\nonumber\\
&-&
\Tr\Bigl[\Bigl(\mat{h}
+\frac{\partial \tilde{F}_\beta^{\hat{W}}}{\partial \mat{\rho}}
\Bigr)\frac{\partial\mat{\rho}}{\partial R_i}\Bigr]
\nonumber\\
\label{eq:forceexpr}
\end{eqnarray}

The first two terms can be identified with a Hellmann-Feynman force,
as they do not depend on the electronic degrees of freedom, namely the
density matrix. Usually the Hellmann-Feynman force is identified with
the electrostatic force on the nucleus, which is of limited practical
value in electronic structure calculations. If the basisset depends
directly on the atomic positions, however, this dependency enters into
derivative of the Hamiltonian and the U-tensor. These terms are often
called Pulay forces\cite{pulay69_molphys17_197}.  Interestingly, the
corresponding contribution from the U-tensor to the Pulay force only
enters through the partial derivative of the Luttinger-Ward
functional, which can be seen from Eq.~\eqref{eq:dmfgreen}.

The terms on the second line vanish for the optimum density matrix,
because the prefactor is directly related to stationary conditions
Eq.~\eqref{eq:minimumrdmf} and
Eq.~\eqref{eq:derivativedensitymatrixfunctional}.  Note, that the
trace of the density matrix is the electron number, which does not
depend on the atomic positions.

It has been found that the Hellmann-Feynman forces, even with the
Pulay forces, are very sensitive to the quality of the convergence,
i.e. on how well the stationary conditions are obeyed. This is due to
the fact that, while the energy only depends in second order on a
deviation from the stationary point, the forces are already sensitive
to the first order. This dependence can be reduced by the terms on the
second line. The most complex approach is to determine the linear
response of the density matrix with respect to atomic positions. This
route is not economical due to the effort for a linear response
calculation.

However, already approximations for the derivative of the density
matrix improve the convergence properties dramatically. This has led
to the development of Andersen's force
theorem\cite{mackintosh80_chapter} for density-functional theory,
where, for example, the electrons within a atom-centered sphere are
displaced with the nucleus.

In the fictitious-Lagrangian approach to \textit{ab initio} molecular
dynamics\cite{car85_prl55_2471}, a completely different route is
taken. Here, electronic degrees of freedom and the atomic positions
are treated on the same footing, and both are propagated according to
Newton's equations of motion. This implies that the forces used in
this approach are \textit{partial} derivatives and not \textit{total}
derivatives with respect to atomic positions. This implies that the
contribution of the electronic response to the force
Eq.~\eqref{eq:forceexpr} \textit{must be} excluded.  In a certain
sense, the electron dynamics takes care of the linear response of the
density matrix. Thus, only the Hellmann-Feynman and Pulay terms need
to be evaluated. The fictitious Lagrangian approach, however, rests on
the existence of a true minium principle, and establishing such a
minimum principle has been one of the major motivations for this work.

\subsection{Accessing the single-particle excitation spectrum}\label{subsec:dynamics}
As outlined at the end of Sec.~\eqref{sec:kbtordmft}, the
density-matrix functional Eq.~\eqref{eq:dmfgreen} together with the
self-consistency equations
\eqref{eq:stationaryG}-\eqref{eq:stationarysigma} describes an
algorithm to calculate a self energy respectively the Green's function
corresponding to a given density matrix.

Let us discuss this feature in some more detail: The first observation
is that the algorithm can be applied to any $N$-representable density
matrix. 
This allows one to employ one approximation for the density-matrix
functional used to optimize the density matrix, and another one to
obtain the single-particle dynamics from the resulting density
matrix. Thus, one can exploit that some approximations are particularly
suited for the total energies, while others have their strengths in
the spectral properties.

For the optimization of the density matrix and atomic positions, one
may apply an algorithm that never resorts to Green's functions and
self energies. Here, one can adopt efficient algorithms that exploit
the minimum principle. The link to the wave-function based formulation of
density-matrix functional theory allows one to use functionals resting
on Eq.~\eqref{eq:dmfphi}, such as the one described
earlier\cite{bloechl11_prb84_205101} or the parametrized density
matrix functionals\cite{mueller84_pl105A_446, baerends01_prl87_133004,
  sharma08_prb78_201103, lathiotakis09_pra79_40501} that refer
neither to the many-particle wave function nor the Green's function.
Another promising approach is the NDE2 approximation described in
Sec.~\ref{sec:nde2}, because it avoids self-consistency of Green's
functions, while allowing us to make close contact to the method used to
extract spectral properties.

Once a density matrix has been obtained, one adopts, in a final step, a
favorite approximation to the Luttinger-Ward functional to extract the
Green's function corresponding to that density matrix.  This is done
following the procedure outlined at the end of
Sec.~\ref{sec:kbtordmft}.  One such approximation to obtain
spectral properties is dynamical mean-field theory, which is obtained
by invoking the local approximation described in
Sec.~\ref{sec:dynamicalmeanfieldtheory}.

It is important to emphasize here that the approximation used to
obtain the density matrix and the approximation for the Luttinger-Ward
functional used for the calculation of spectral properties need not be
related in any way. For example, one can use density-functional theory
to calculate a density matrix, and a Hartree-Fock functional for the
Green's function, thereby producing genuine quasi-particle
bandstructures without resorting to Koopman's
theorem.\cite{koopmans34_physica1_104,szabo89_book} Alternatively, one can
employ the Luttinger-Ward functional of the dynamical mean-field
theory at this point, which produces an approximation that at first
sight seems to be the well-established
density-functional-theory + dynamical-mean-field theory (DFT+DMFT)
approach.

There are, however, important differences between our approach and the
conventional setup of DFT+DMFT.  In the DFT+DMFT hybrid approaches,
the non-interacting Hamiltonian is extracted from a non-interacting
electron gas with the same electron density as the interacting
electron gas. While the Kohn-Sham bands provide a surprisingly good
description of quasi-particle bandstructures, there is no apparent
conceptual connection with the non-interacting part of the Hamiltonian
entering the dynamical mean-field theory. Density-matrix functional
theory, on the contrary, is at an advantage, because it uses a
one-particle Hamiltonian $\mat{h'}$ that is linked to the interacting
electrons: It is the Lagrange multiplier obtained via
Eq.~\eqref{eq:stationarysigma} and Eq.~\eqref{eq:stationaryhprime}
from the density matrix constraint for the density matrix of the
interacting electron gas. As a result, the spectra obtained in the
final step from the rDMFT method can be interpreted directly as
physical excitation spectra, without the well-known interpretation
problems of Kohn-Sham spectra.


If such a hybrid scheme is built into a DFT environment, there is a
fairly well defined expression for the double counting term within the
DFT+rDMFT approach.\cite{bloechl11_prb84_205101} Furthermore, the
double-counting term, only enters the first step, namely the
determination of the density matrix. For the final step of
  calculating the Green's function from a given density matrix,
  double-counting corrections are not even required.

\section{Conclusions}
A link between Green's-function diagrammatic techniques and the
reduced density-matrix functional theory has been established by a
re-formulation of the density-matrix functional involving a
constrained search over single-particle Green's functions.  This
constrained search is equivalent to solving the corresponding
stationarity conditions and can be performed in practice by setting up
a self-consistency scheme.  Diagrammatic weak-coupling approaches as
well as the non-perturbative dynamical mean-field theory are obtained
as specific approximations of the density-matrix functional in this
way.

The reformulation of diagrammatic approximations within the context of
density-matrix functional theory is of great importance as the latter
provides a true minimum principle.  This is opposed to "dynamical"
functionals, i.e. functionals of the Green's-function or the
self-energy which are generally not convex.  Attempts to enforce the
convexity by modifying the Kadanoff-Baym functional or other dynamical
functionals would require  modifying their global free-energy landscape
in a qualitative manner and will thus most likely also distort the
physically relevant properties of the functionals.

As the one-particle density matrix is obtained by a
Matsubara-frequency summation of the single-particle Green's function,
the density-matrix functional theory can be seen as an approach where
a certain hypersurface in the space of electronic degrees of freedom
is selected.  This hypersurface picks those "static" degrees of
freedom that result in a convex free-energy landscape while the free
energy, or the grand potential, shows the "wrong" curvature along
"dynamical" coordinates orthogonal to this hypersurface.

Clearly, convexity can be shown rigorously for the \textit{exact}
density-matrix functional only and is not necessarily true for
approximations.  Note that this is the same situation as in DFT: Many
rigorous properties of exact density-functional theory do not carry
over to approximate density functionals, e.g.\ to the LDA functional.

It is obvious, however, that a convex free-energy landscape is
obtained much more easily in an approximation to an exact functional
that is convex already.  That is, one would expect that reliable
approximations inherit the minimum property from the exact rDMFT
theory.  Or stated differently, one would reject diagrammatic
approximations which do qualitatively change the topology of the rDMFT
free-energy surface or even the local free-energy surface close to the
physical point.

The case of dynamical mean-field theory is particularly interesting.
Numerical work related to the question of convexity of the
density-matrix functional derived from dynamical mean-field theory is
in progress.  Another exciting line of further development is to
completely avoid the computation of the spectral function within
dynamical mean-field theory, i.e., to set up a self-consistent rDMFT
scheme involving static ground-state or thermal properties only, such
as energies, reduced density matrices or forces.  Here, one would
profit from a formulation based on a minimum principle during the
search for self-consistency while spectral properties are still
accessible in a final post-processing step as outlined in
Sec.~\ref{subsec:dynamics}.

The price one has to pay is that one now needs to adjust the full
one-particle Hamiltonian $\mat{h}'$ together with the single-particle
self energy to fulfill the density-matrix constraint. The complexity
of this step is, however, not higher than that of a
dynamical-mean-field-theory self-consistency
without charge optimization.


\begin{acknowledgements}
Financial support by the Deutsche
Forschungsgemeinschaft through FOR 1346 is gratefully acknowledged.
\end{acknowledgements}
\appendix
\section{Convexity of the density-matrix functional}
\label{app:dmfconvexity}
Here we show that the density-matrix functional defined by
Eq.~\eqref{eq:dmfphi} as
\begin{widetext}
\begin{eqnarray}
F[\mat{\rho}]&=&
\min_{P_j\ge0,|\Phi_j\rangle}\stat_{\tilde{\mat{h}},\mat{\Lambda},\lambda}
\biggl\lbrace \sum_jP_j\langle\Phi_j|\hat{W}|\Phi_j\rangle
+\frac{1}{\beta}\sum_j P_j\ln[P_j]
\nonumber\\ 
&&-\sum_{i,j}\Lambda_{i,j}\Bigl(\langle\Phi_j|\Phi_i\rangle-\delta_{j,i}\Bigr)
-\lambda\Bigl(\sum_j P_j-1\Bigr) +\sum_{a,b}\tilde{h}_{a,b}\Bigl(
\sum_j
P_j\langle\Phi_j|\hat{c}^\dagger_a\hat{c}_b|\Phi_j\rangle-\rho_{b,a}
\Bigr) \biggr\rbrace
\end{eqnarray}
\label{eq:variantofdmfphi}
\end{widetext}
is convex. That is
\begin{eqnarray}
F[(1-\lambda)\mat{\rho}^A+\lambda\mat{\rho}^B]\le
(1-\lambda)F[\mat{\rho}^A]+\lambda F[\mat{\rho}^B]\;.
\label{eq:convexcondition}
\end{eqnarray}

The proof is facilitated by the fact that the one-particle density
matrix and all expectation values depend only linearly on the
many-particle density matrix.

Before we start, let us introduce a few quantities: The constrained
search for the reduced one-particle density matrix $\mat{\rho}^A$
leads to an optimum ensemble $\{|\Phi_j^A\rangle, P^A_j\}$
characterized by the many-particle density matrix
$\hat{\Gamma}^A=\sum_j |\Phi^A_j\rangle
P^A_j\langle\Phi^A_j|$. Analogously, the many-particle density matrix
$\hat{\Gamma}^B$ for the ensemble $\{|\Phi_j^B\rangle, P^B_j\}$ is
connected to the one-particle reduced density matrix $\mat{\rho}^B$.
We define the $\lambda$-dependent many-particle density matrix 
\begin{eqnarray}
\hat{\Gamma}(\lambda):=(1-\lambda)\hat{\Gamma}^A+\lambda\hat{\Gamma}^B
\;.
\label{eq:gammaoflambda}
\end{eqnarray}
which linearly connects the two ensembles.  Diagonalization of
$\hat{\Gamma}(\lambda)$ yields the many-particle wave functions
$|\Phi_j(\lambda)\rangle$ and the eigenvalues $P_j(\lambda)$.

At first, we show that the many-particle density matrix
$\hat{\Gamma}(\lambda)$ obeys the constraints required for the
optimization. These constraints are (1) the orthonormality of the
many-particle wave functions, (2) the normalization constraint of the
probabilities, (3) the requirements that the all probabilities are
positive, and (4) the density-matrix constraint. Finally, (5), the ensemble
wave functions must be antisymmetric under particle exchange.

The orthonormality of the many-particle wave functions
$|\Phi_j(\lambda)\rangle$ in the ensemble follows directly from the
fact that they are eigenstates of a Hermitian operator
$\hat{\Gamma}(\lambda)$. To be precise, the eigenstates can be chosen
to be orthonormal.

The normalization of the probabilities, i.e., $\sum_j P_j=1$, can be
expressed as $\Tr[\hat{\Gamma}(\lambda)]=1$.
It is easily shown that
\begin{eqnarray}
\Tr[\hat{\Gamma}(\lambda)]
&=&\Tr[(1-\lambda)\hat{\Gamma}^A+\lambda\hat{\Gamma}^B]
\nonumber\\
&=&(1-\lambda)\Tr[\hat{\Gamma}^A]+\lambda\Tr[\hat{\Gamma}^B]=1\;,
\end{eqnarray}
which proves that $\hat{\Gamma}(\lambda)$ obeys the normalization
constraint.

The positive definiteness of $\hat{\Gamma}(\lambda)$,
i.e. $P_j(\lambda)\ge0$, follows directly from the positive
definiteness of $\hat{\Gamma}^A$ and $\hat{\Gamma}^B$. We need to show
that for any many-particle state $|\Psi\rangle$ the expectation value
$\langle\Psi|\hat{\Gamma}(\lambda)|\Psi\rangle$ is non-negative.
We obtain
\begin{eqnarray}
\langle\Psi|\hat{\Gamma}(\lambda)|\Psi\rangle
=(1-\lambda)\langle\Psi|\hat{\Gamma}^A|\Psi\rangle
+\lambda \langle\Psi|\hat{\Gamma}^B|\Psi\rangle\ge 0
\;,
\nonumber\\
\end{eqnarray}
if $\langle\Psi|\hat{\Gamma}^A|\Psi\rangle\ge0$ and
$\langle\Psi|\hat{\Gamma}^B|\Psi\rangle\ge0$, and if
$0\le\lambda\le1$.

Next, we need to show  that the density-matrix constraint 
\begin{eqnarray}
\rho_{\beta,\alpha}(\lambda)=\Tr[\hat{\Gamma}(\lambda)
\hat{c}^\dagger_\alpha\hat{c}_\beta]
\end{eqnarray}
is obeyed. It is verified as follows
\begin{eqnarray}
\Tr[\hat{\Gamma}(\lambda)
\hat{c}^\dagger_\alpha\hat{c}_\beta]
&=&\Tr\Bigl[\Bigl((1-\lambda)\hat{\Gamma}^A+\lambda\hat{\Gamma}^B\Bigr)
\hat{c}^\dagger_\alpha\hat{c}_\beta\Bigr]
\nonumber\\
&=&
(1-\lambda)\Tr\Bigl[\hat{\Gamma}^A\hat{c}^\dagger_\alpha\hat{c}_\beta\Bigr]
+\lambda\Tr\Bigl[\hat{\Gamma}^B\hat{c}^\dagger_\alpha\hat{c}_\beta\Bigr]
\nonumber\\
&=&
(1-\lambda)\rho^A_{\beta,\alpha}
+\lambda\rho^B_{\beta,\alpha}
=
\rho_{\beta,\alpha}(\lambda)\;.
\end{eqnarray}

Finally, the antisymmetry also carries over from the end-points due to
the linearity of Eq.~\eqref{eq:gammaoflambda}.

As all the constraints are obeyed by $\hat{\Gamma}(\lambda)$, we
obtain an upper bound for the density-matrix functional by evaluating
the energy contributions of \eqref{eq:variantofdmfphi} with
$\hat{\Gamma}(\lambda)$, i.e.
\begin{eqnarray}
F[\mat{\rho}(\lambda)]\le
Tr[\hat{\Gamma}(\lambda)\hat{W}]
+\frac{1}{\beta}
Tr[\hat{\Gamma}(\lambda)\ln\Bigl(\hat{\Gamma}(\lambda)\Bigr)\Bigr]
\;.
\end{eqnarray}

The interaction energy depends linearly on the many-particle density
matrix, so that
\begin{eqnarray}
\Tr[\hat{\Gamma}(\lambda)\hat{W}]=
(1-\lambda)\Tr[\hat{\Gamma}^A\hat{W}]+\lambda\Tr[\hat{\Gamma}^B\hat{W}]\;,
\end{eqnarray}
and the entropy $-k_B\Tr[\hat{\Gamma}\ln(\hat{\Gamma})]$ is
concave\cite{wehrl78_rmp50_221}.

With 
\begin{eqnarray}
F[\mat{\rho}^A]=
Tr[\hat{\Gamma}^A\hat{W}]
+\frac{1}{\beta}Tr[\hat{\Gamma}^A\ln(\hat{\Gamma}^A)\Bigr]
\\
F[\mat{\rho}^B]=
Tr[\hat{\Gamma}^B\hat{W}]
+\frac{1}{\beta}Tr[\hat{\Gamma}^B\ln(\hat{\Gamma}^B)\Bigr]\;,
\end{eqnarray}
we obtain
\begin{eqnarray}
F[\mat{\rho}(\lambda)]\le
(1-\lambda)F[\mat{\rho}^A]+\lambda F[\mat{\rho}^B]\;,
\end{eqnarray}
which is equivalent to Eq.~\eqref{eq:convexcondition}. This concludes
the proof that the density-matrix functional is convex.

\section{Invariance of the Kadanoff-Baym functional}
\label{app:invariance}
Here, we prove Eq.~\eqref{eq:transformationpsikb}.
We introduce  
\begin{eqnarray}
&&\hspace{-1cm}
Y(\mat{\Delta}):=\Psi^{KB}_{\beta,\mu}[\mat{G},\mat{\Sigma}+\mat{\Delta}
,\mat{h}-\mat{\Delta},\hat{W}] 
\nonumber\\
&+&\frac{1}{\beta}\sum_\nu\e{i\beta\hbar\omega_\nu0^+}\Tr[\mat{G}\mat{\Delta}]
-\Psi^{KB}_{\beta,\mu}[\mat{G},\mat{\Sigma},\mat{h},\hat{W}] 
\;,
\end{eqnarray}
which vanishes, when Eq.~\eqref{eq:transformationpsikb} is valid.

With Eq.~\eqref{eq:kadanoffbaym2}, we obtain
\begin{eqnarray}
Y(\mat{\Delta})
&=&
\frac{1}{\beta}\sum_\nu\e{i\beta\hbar\omega_\nu0^+}
\Tr\ln\left(\frac{(i\hbar\omega_\nu+\mu)\mat{1}-\mat{h}+\mat{\Delta}}
{(i\hbar\omega_\nu+\mu)\mat{1}-\mat{h}}\right)
\nonumber\\
&-&\frac{1}{\beta}\Tr\ln\left(
\frac{\mat{1}+\e{-\beta(\mat{h}-\mat{\Delta}-\mu\mat{1})}}
{\mat{1}+\e{-\beta(\mat{h}-\mu\mat{1})}}\right)
\end{eqnarray}
It is easily seen that $Y(\mat{0})=0$. In order to prove that
$\mat{Y}(\mat{\Delta})=0$ for all arguments, we need to show that its
derivative vanishes. We obtain
\begin{eqnarray}
\frac{\partial Y}{\partial\mat{\Delta}}
&=&\frac{1}{\beta}\sum_\nu\e{i\beta\hbar\omega_\nu0^+}
\frac{1}{(i\hbar\omega_\nu+\mu)\mat{1}-\mat{h}+\mat{\Delta}}
\nonumber\\
&-&\frac{1}{1+\e{\beta(\mat{h}-\mat{\Delta}-\mu\mat{1})}}=0
\;,
\end{eqnarray}
which is verified using the elementary Matsubara
sum\cite{fetter71_book}
$\frac{1}{\beta}\sum_\nu\e{i\beta\hbar\omega_\nu0^+}
\frac{1}{i\hbar\omega_\nu-\epsilon}=\frac{1}{1+\e{\beta\epsilon}}$.
This concludes the proof.


\section{Functional derivative of the density-matrix functional}
\label{app:dfdrho}
Here, we show the explicit derivation of the derivative
Eq.~\eqref{eq:derivativedensitymatrixfunctional} of the density-matrix
functional Eq.~\eqref{eq:dmfgreen} and the minimum condition
Eq.~\eqref{eq:minimumrdmf} for the grand potential

We begin with Eq.~\eqref{eq:dmfgreen}
\begin{eqnarray}
\tilde{F}^{\hat{W}}_\beta[\mat{\rho}]
&=&
\frac{1}{\beta}\Tr\Bigl[
\mat{\rho}\ln(\mat{\rho})+(\mat{1}-\mat{\rho})\ln(\mat{1}-\mat{\rho})\Bigr]
\nonumber\\
&+&
\stat_{\mat{h}'}\stat_{\mat{G},\mat{\Sigma}}
\biggl\lbrace
\Phi^{LW}_\beta[\mat{G},\hat{W}]
\nonumber\\
&-&\frac{1}{\beta}\sum_\nu\Tr\Bigl\lbrace
\ln\Bigl[
\mat{1}-\matGrho
\Bigl(\mat{h}'+\mat{\Sigma}-\mathrho\Bigr)
\Bigr]
\nonumber\\
&+&(\mat{h}'+\mat{\Sigma}-\mathrho)\mat{G}
-
\Bigl[\mat{G}-\matGrho\Bigr]
\Bigl(\mat{h}'-\mathrho\Bigr)\Bigr\rbrace
\biggr\rbrace
\nonumber\\
\end{eqnarray}
where we used $\matGrho(i\omega_\nu)$ defined earlier in
Eq.~\eqref{eq:defgrho}. 

We consider the dependencies via $\mathrho$ and $\matGrho$,
while we exploit the stationary condition with respect to
$\mat{G},\mat{\Sigma}$ and $\mat{h}'$.
\begin{eqnarray}
\frac{\partial\tilde{F}_\beta}{\partial\mat{\rho}}
&=&
\frac{1}{\beta}\ln\left(\frac{\mat{\rho}}{\mat{1}-\mat{\rho}}\right)
\nonumber\\
&-&\frac{1}{\beta}\sum_\nu\Tr\biggl\lbrace
\Bigl[
\mat{1}-\matGrho
\Bigl(\mat{h}'+\mat{\Sigma}-\mathrho\Bigr)
\Bigr]^{-1}
\nonumber\\
&&\times
\underbrace{
\Bigl[
-\matGrho\frac{\partial\mathrho}{\partial\mat{\rho}}
\matGrho
\Bigl(\mat{h}'+\mat{\Sigma}-\mathrho\Bigr)
+\matGrho\frac{\partial\mathrho}{\partial\mat{\rho}}
\Bigr]}
_{\matGrho\frac{\partial\mathrho}{\partial\mat{\rho}}
\Bigl[1-\matGrho
\Bigl(\mat{h}'+\mat{\Sigma}-\mathrho\Bigr)\Bigr]}
\nonumber\\
&-&\frac{\partial\mathrho}{\partial\mat{\rho}}
\mat{G}
+
\frac{\partial\matGrho}{\partial\mat{\rho}}
\Bigl(\mat{h}'-\mathrho\Bigr)
+
\Bigl[\mat{G}-\matGrho\Bigr]
\frac{\partial\mathrho}{\partial\mat{\rho}}
\biggr\rbrace
\nonumber\\
&=&-\mathrho+\mu\mat{1}
-\frac{1}{\beta}\sum_\nu\Tr\biggl\lbrace
\matGrho\frac{\partial\mathrho}{\partial\mat{\rho}}
\nonumber\\
&-&
\frac{\partial\mathrho}{\partial\mat{\rho}}
\mat{G}
+\frac{\partial\matGrho}{\partial\mat{\rho}}
\Bigl(\mat{h}'-\mathrho\Bigr)
+
\Bigl[
\mat{G}
-
\matGrho
\Bigr]
\frac{\partial\mathrho}{\partial\mat{\rho}}
\biggr\rbrace
\nonumber\\
&=&-\mathrho+\mu\mat{1}-\frac{1}{\beta}\sum_\nu\Tr\biggl\lbrace
\frac{\partial\matGrho}{\partial\mat{\rho}}
\Bigl(\mat{h}'-\mathrho\Bigr)
\biggr\rbrace
\nonumber\\
&=&-\mathrho+\mu\mat{1}-\Tr\biggl\lbrace
\Bigl(\mat{h}'-\mathrho\Bigr)
\frac{\partial}{\partial\mat{\rho}}
\Bigl(
\frac{1}{\beta}\sum_\nu\matGrho\Bigr)
\biggr\rbrace
\nonumber\\
&=&-(\mat{h}'-\mu\mat{1})
\nonumber\\
\label{eq:dfdrho1}
\end{eqnarray}

This is the desired expression
Eq.~\eqref{eq:derivativedensitymatrixfunctional} for the derivative of
the density-matrix functional.

Now, we turn to the stationary condition Eq.~\eqref{eq:minimumrdmf} for
the grand potential 
\begin{eqnarray}
\Omega_{\beta,\mu}(\hat{h}+\hat{W})&=&\min_{|\psi_n\rangle,f_n\in[0,1]}\stat_{\mat{\Lambda}}\biggl\lbrace
\sum_n f_n\langle\psi_n|\hat{h}|\psi_n\rangle
\nonumber\\
&+&F^{\hat{W}}_{\beta}\Bigl[\sum_n|\psi_n\rangle f_n\langle\psi_n|\Bigr]
-\mu\sum_n f_n
\nonumber\\
&-&\sum_{m,n}\Lambda_{m,n}\Bigl(\langle\psi_n|\psi_m\rangle-\delta_{n,m}\Bigr)
\biggr\rbrace
\label{eq:gcpdm1}
\end{eqnarray}
specified in Eq.~\eqref{eq:gcpdm2}.

The minimum conditions for Eq.~\eqref{eq:gcpdm1} with respect to the
natural orbitals and occupations are
\begin{eqnarray}
\Bigl(\hat{h}+\frac{\partial F^{\hat{W}}_\beta}{\partial\hat{\rho}}\Bigr)
|\psi_n\rangle-\sum_m|\psi_m\rangle\Lambda_{m,n}\frac{1}{f_n}=0
\label{eq:domegadpsi}
\\
\langle\psi_n|\hat{h}+\frac{\partial F^{\hat{W}}_\beta}{\partial\hat{\rho}}
|\psi_n\rangle-\mu
=0
\label{eq:domegadf}
\end{eqnarray}

From the first equation, Eq.~\eqref{eq:domegadpsi}, we obtain,
exploiting orthonormality of the natural orbitals $|\psi_n\rangle$,
\begin{eqnarray}
\langle\psi_m|
\Bigl(\hat{h}+\frac{\partial F^{\hat{W}}_\beta}{\partial\hat{\rho}}\Bigr)
|\psi_n\rangle f_n=\Lambda_{m,n}
\label{eq:domegadbra}
\end{eqnarray}
From the variation with respect to the ket $\langle\psi_n|$, we obtain
an equation analogous to Eq.~\eqref{eq:domegadbra}
\begin{eqnarray}
f_m\langle\psi_m|
\Bigl(\hat{h}+\frac{\partial F^{\hat{W}}_\beta}{\partial\hat{\rho}}\Bigr)
|\psi_n\rangle =\Lambda_{m,n}
\label{eq:domegadket}
\end{eqnarray}
The two equations Eq.~\eqref{eq:domegadbra} and Eq.~\eqref{eq:domegadket}
can be combined into
\begin{eqnarray}
\langle\psi_m|
\Bigl(\hat{h}+\frac{\partial F^{\hat{W}}_\beta}{\partial\hat{\rho}}\Bigr)
|\psi_n\rangle(f_m-f_n)=0
\label{eq:domegadbracket}
\end{eqnarray}

As long as the occupations are all different,
Eqs.~\eqref{eq:domegadbracket}
and
Eq.~\eqref{eq:domegadf} can be combined into the condition
\begin{eqnarray}
\frac{\partial F^{\hat{W}}_\beta}{\partial\hat{\rho}}
=-\hat{h}+\mu\hat{1}
\label{eq:equildensitymatrixfunctional}
\end{eqnarray}

Combining Eq.~\eqref{eq:dfdrho1} and
Eq.~\eqref{eq:equildensitymatrixfunctional}, the minimum condition
for Eq.~\eqref{eq:gcpdm1} with respect to natural orbitals and
occupations has the form
\begin{eqnarray}
\hat{h}=\hat{h}'
\end{eqnarray}
where $\mat{h}$ is the external potential and kinetic energy, while
$\mat{h}'$ is part of the Lagrange multiplier from the density-matrix
constraint in the density-matrix functional. 

\section{Adiabatic connection at finite temperature}
\label{app:adiabatic}
The grand potential in density-functional theory is 
\begin{eqnarray}
\Omega_{\beta,\mu}[v_{ext}]&=&
\min_{n(\vec{r})}\Bigl\lbrace G^{\hat{W}}_{\beta}[n]
+\int d^3r\; n(\vec{r})v_{ext}(\vec{r})\Bigr\rbrace
\end{eqnarray}
where $G^{\lambda\hat{W}}_{\beta}[n]$ is the universal density
functional for a $\lambda$-scaled interaction $\lambda\hat{W}$
\begin{eqnarray}
G^{\lambda\hat{W}}_\beta[n]&:=&\min_{\{|\Phi_j\rangle,P_j\}\in M[n]}
\Bigl\lbrace\sum_jP_j\langle\Phi_j|\hat{T}+\lambda\hat{W}|\Phi_j\rangle
\nonumber\\
&+&\frac{1}{\beta}\sum_j P_j\ln(P_j)
\Bigr\rbrace
\nonumber\\
&=&
\sum_j\bar{P}_j(\lambda)\langle\bar{\Phi}_j(\lambda)|
\hat{T}+\lambda\hat{W}|\bar{\Phi}_j(\lambda)\rangle
\nonumber\\
&+&\frac{1}{\beta}\sum_j \bar{P}_j(\lambda)\ln(\bar{P}_j(\lambda))
\;,
\end{eqnarray}
where $\hat{T}$ is the kinetic energy operator.  In the first line,
$M[n]$ is the set of all fermionic many-particle ensembles that
produce the density $n(\vec{r})$. The many-particle ensembles are
characterized by antisymmetric and orthonormal many-particle wave
functions and non-negative probabilities that add up to one. The
$\lambda$-dependent quantities $|\bar{\Phi}_j(\lambda)\rangle$ and
$\bar{P}_j(\lambda)$ are those obtained by satisfying the minimum
conditions with the scaled interaction $\lambda\hat{W}$.

Now, we follow the ensemble for a given density
adiabatically from the non-interacting to the fully interacting
case. The functional at full interaction is obtained as
interaction-strength integral of the derivative. The latter can be
simplified by exploiting the Hellmann-Feynman theorem, that is, by
exploiting that the derivatives with respect to wave functions and
probabilities vanish.
\begin{eqnarray}
G^{\hat{W}}_\beta[n]&=&
G^{\hat{0}}_\beta[n]+\int_0^1d\lambda\;
\frac{dG^{\lambda\hat{W}}_\beta[n]}{d\lambda}
\nonumber\\
&=&K_\beta[n]
+\int_0^1d\lambda\;
\sum_j \bar{P}_j(\lambda)\langle
\bar{\Phi}_j(\lambda)|\hat{W}|\bar{\Phi}_j(\lambda)\rangle
\nonumber\\
&=&K_\beta[n]+E_H[n]+\int_0^1d\lambda\; U_{xc}^{\lambda,\beta}[n]
\end{eqnarray}

The intrinsic energy $K_\beta[n]$ of non-interacting electrons
is calculated as a minimization over Kohn-Sham wave functions
$|\psi^{KS}_n\rangle$ and their occupations $f_n^{KS}$.
\begin{eqnarray}
K_\beta[n]
&=&\min_{f^{KS}_n,|\psi^{KS}_n\rangle}\stat_{v_{eff},\mat{\Lambda}}
\sum_n f^{KS}_n\langle\psi^{KS}_n|\frac{\hat{\vec{p}}^2}{2m_e}|\psi^{KS}_n\rangle
\nonumber\\
&+&\frac{1}{\beta}
\sum_n\Bigl[f^{KS}_n\ln(f^{KS}_n)+(1-f^{KS}_n)\ln(1-f^{KS}_n)\Bigr]
\nonumber\\
&+&\int d^3r\; v_{eff}(\vec{r})
\Bigl[\sum_{n}f^{KS}_n|\psi^{KS}_n(\vec{r})|^2-n(\vec{r})\Bigr]
\nonumber\\
&-&\sum_{m,n}\Lambda_{m,n}
\Bigl(\langle\psi^{KS}_n|\psi^{KS}_m\rangle-\delta_{n,m}\Bigr)
\end{eqnarray}
The effective potential $v_{eff}(\vec{r})$ is, like $\mat{\Lambda}$, a
Lagrange multiplier. 

$E_{H}$ is the Hartree energy 
\begin{eqnarray}
E_{H}[n]:=\frac{1}{2}\int d^3r\int d^3r'\;
\frac{e^2n(\vec{r})n(\vec{r'})}{4\pi\epsilon_0|\vec{r}-\vec{r'}|}
\end{eqnarray}
and
\begin{eqnarray}
U^{\lambda,\beta}_{xc}[n]=\int d^3r\;n(\vec{r})
\left\lbrace\frac{1}{2}
\int d^3r'
\frac{e^2h_{\beta,\lambda}(\vec{r},\vec{r'})}{4\pi\epsilon_0|\vec{r}-\vec{r'}|}
\right\rbrace
\nonumber\\
\label{eq:Uxc}
\end{eqnarray}
is the electrostatic interaction energy of the electrons with their
respective exchange-correlation hole
$h_{\beta,\lambda}(\vec{r},\vec{r'})$. The latter is obtained at the
specified temperature and scaled interaction strength. The hole
function $h_{\beta,\lambda}(\vec{r},\vec{r'})$ is related to the
two-particle density $n^{(2)}(\vec{r},\vec{r'})$ via
$n^{(2)}(\vec{r},\vec{r'})=n(\vec{r})
\Bigl(n(\vec{r'})+h_{\beta,\lambda}(\vec{r},\vec{r'})\Bigr)$.  Without
interaction, Eq.~\eqref{eq:Uxc} provides the exchange energy
$U^{0,\beta}_{xc}[n]$.

Thus, the exchange correlation energy $E_{xc}[n]$ can be constructed,
like in the zero-temperature limit, as an interaction-strength average
of a purely electrostatic term.
\begin{eqnarray}
E_{xc}[n]=\int_0^1d\lambda\;U_{xc,\beta}^{\lambda}[n]
\end{eqnarray}
The exchange correlation hole, however, is temperature dependent.  Via
the Hellmann-Feynman theorem, it contains contributions from kinetic
energy and entropy term.  The main contribution to kinetic and entropy
term is contained in $K_\beta[n]$ which considers both contributions
of the non-interacting electron gas.

\bibliography{all}
\end{document}